\documentclass[iop]{emulateapj}
\pdfoutput=1
\usepackage{graphicx, subfigure}
\usepackage{morefloats}
\usepackage{amsmath}
\usepackage{amssymb}
\usepackage{mathrsfs,amssymb}
\usepackage{url}
\usepackage{color}

\usepackage[percent]{overpic}

\usepackage{rotating}

\usepackage{floatrow}

\usepackage{natbib}

\usepackage{letltxmacro}

\LetLtxMacro{\ORIcitep}{\citep}
\DeclareRobustCommand{\citep}{\leavevmode\unskip~\ORIcitep}

\begin{document}

\title{Central Engine Memory of Gamma-Ray Bursts and Soft Gamma-Ray Repeaters}

%

\author{Bin-Bin Zhang\altaffilmark{1,2}, Bing Zhang\altaffilmark{3},Alberto J. Castro-Tirado\altaffilmark{1}}

\altaffiltext{*}{Contact Email: zhang.grb@gmail.com}

\affil{$^{1}$Instituto de Astrof\'isica de Andaluc\'a (IAA-CSIC), P.O. Box 03004, E-18080 Granada, Spain}
\affil{$^{2}$Center for Space Plasma and Aeronomic Research (CSPAR), University
of Alabama in Huntsville, Huntsville, AL 35899, USA} 
\affil{$^{3}$Department of Physics and Astronomy, University of Nevada, Las Vegas, NV 89154, USA}


%
%

\begin{abstract}
Gamma-ray Bursts (GRBs) are bursts of $\gamma$-rays generated from relativistic jets launched from catastrophic events such as massive star core collapse or binary compact star coalescence.
Previous studies suggested that GRB emission is erratic, with no noticeable memory in the central engine. Here we report a discovery that similar light curve patterns exist within individual bursts for at least some GRBs. Applying the Dynamic Time Warping (DTW) method, we show that similarity of light curve patterns between pulses of a single burst or between the light curves of a GRB and its X-ray flare can be identified. This suggests that the central engine of at least some GRBs carries ``memory'' of its activities. 
We also show that the same technique can identify memory-like emission episodes in the flaring emission in Soft Gamma-Ray Repeaters (SGRs), which are believed to be Galactic, highly magnetized neutron stars named magnetars. {Such a phenomenon challenges the standard black hole central engine models for GRBs, and suggest a common physical mechanism behind GRBs and SGRs, which points towards a magnetar central engine of GRBs.}
 \end{abstract}

\section{Introduction}

\label{sect:main}

In astronomy, light curves usually carry the information about the central objects. In stably rotating neutron star systems (e.g. radio pulsars and X-ray pulsars), strict periodicity is detected, which is related to the rotation period of the neutron star. In black hole binaries or some active galactic nuclei (AGNs), quasi-periodic oscillation (QPO) signals are detected in certain frequency ranges, which probe the quasi-periodic properties in the accreting systems. For transient objects such as GRBs\citep[e.g.,][]{meszaros06,gehrels09,kumarzhang15} and SGRs\citep[e.g,][]{kouveliotou98,mereghetti15}, on the other hand, the light curves usually track the history of the central engine activities. Time sequence analyses of GRB light curves using power density spectral analysis and fast Fourier transform \citep{beloborodov00,guidorzi12,hakkila14,baldeschi15} suggested that these light curves are ``featureless''. Using a stepwise filter correlation method, \cite{gao12} revealed the existence of superpositions of slow and fast components in GRB light curves, but no repetitive feature of GRB emission was identified. It is believed that due to the erratic activities of the central engine (e.g. unsteady accretion onto a black hole), one would not expect a repeatable pattern in GRB light curves \citep{meszaros06,kumarzhang15}. In other words, the GRB central engine is supposed to be ``memory-less''. Similarly, the SGR flares mark the erratic magnetic reconnection processes in the crusts or magnetospheres of the magnetars \citep{thompson95}. No regular repeatable patterns are expected.

On May 8, 2014, GRB 140508A was detected by Fermi/GBM\citep{2014GCN..16224...1Y}, which exhibits 6 visually dramatically similar episodes in its light curve (Figure 1). Multi-pulse GRBs have been commonly observed\citep{2011A&A...528A..15G}, but the one with such a peculiar signature has not been noticed before. Since power density spectral analyses usually do not catch these features, we therefore apply other analysis methods as an effort to confirm the visual impression of repeatable behavior in this burst. { Since GRB light curves are defined by many factors, including the central engine activity, dynamical evolution (acceleration, coasting, or deceleration) during the emission phase, and the viewing angle, the light curves in different epochs may have been distorted in the time domain even if different episodes have intrinsic similarities. Traditional methods, such as the Bayesian Adaptive Regression Splines (BARS) method introduced by KASS (http://www.stat.cmu.edu/~kass/bars/), are not ideal to correct for these distortions.} 
After testing some methods, we find that the dynamic time warping \citep[DTW][]{Sakoe78} technique is well suited to address this problem. DTW is a robust technique to search for similar time series signals by automatically coping with time deformation and varying speed associated with time-dependent data, and has been widely used in diverse fields\citep{Keogh05}. This method calculates an optimal match between two time sequences by minimizing the distance-like quantity. By effectively stretching or squeezing the time sequences, this method can account for non-linear variation in the time dimension by non-linearly ``warping" the time sequence to achieve a similarity match with another time sequence. 

In this letter, we apply the DTW method to the light curves of GRBs and SGRs and try to identify the similar patterns of the central engine activities of these objects. Our method is detailed in Section 2, results are presented in Section 3. A brief summary and discussion on astrophysical implications are presented in Section 4.

\section{Method}

{ The basic concept of the DTW method and a toy example to show how it works are presented in the Appendix.} Practically, the following procedure can be used to search for a similarity pattern in a light curve. (I) Pattern selection: One may select a {\em query pattern} in the light curve (denoted as $X=(x_1,...,x_N)$) to be searched. 
(II) Reference selection: By cutting the pattern off the light curve, the rest of the light curve is regarded as the {\rm reference} (denoted as $Y=(y_1,...,y_M)$), which is the time series where patterns similar to the pattern $X$ may be searched. (III) Search for pattern in the reference: The open-begin--end DTW (OBE-DTW) technique \citep{Paolo08} is performed in this step. The goal of a match is to minimize the average accumulated distortion, $D$, between the warped time series X and Y$_{p-q}$, i.e.
\begin{eqnarray}
&& Y^{(p,q)}=(y_p,...,y_q) \nonumber \\
&& D_{OBE}=\min\limits_{1\leq p \le q \leq M} D (X, Y^{(p,q)}) 
\end{eqnarray}
where $D (X, Y^{(p,q)})= \Sigma d(\phi_x(k),\phi_y(k)) m_\phi(k)/M_\phi$, and $k=1,...,T$, with $T$ being the allowable absolute time deviation between the two matched elements. The distance function is defined as $d(i,j)=|x_i-y_i|$. $M_\phi$ is the normalization constant so that the accumulated distortions are comparable along different paths\citep{Toni09}, and $m_\phi(k)$ is a weighting coefficient. The warping curve $\phi(k)$ is defined as:
\begin{equation}
\phi(k)= (\phi_x(k),\phi_y(k))
\end{equation}
with $\phi_x(k)\in \{1,...,N\}$ and $\phi_x(k)\in \{p,...,q\}$. The DTW algorithm requires certain constraints on $\phi$, e.g., monotonicity is imposed to preserve their time ordering and to avoid meaningless loops: $\phi_x(k+1)\ge \phi_x(k)$, $\phi_y(k+1)\ge \phi_y(k)$, and $p$ and $q$ are determined during minimizing the cumulative distance in Equation (1). After the search, one may identify a ``successful match" (criteria defined below) or a ``rejected match". 
(IV) After finding the match, the matched segment is cut off from the reference, and the step (III) is repeated to search for more matches.

For each match, we measure the correlation between the normalized pattern light curve and the normalized de-warped reference light curve. We define a successful match based on the following criteria:
(a) The Pearson's correlation coefficient is $r >0.8$, which indicates a ``very strong" correlation\citep{Evans96}. (b) The Spearman rank correlation significance level, $p$ is less than 0.003 (3-$\sigma$ level). We note that such criteria, even though carefully chosen based on statistical methods, have not been applied to the DTW applications before. Strictly speaking, the threshold criteria should be tested through experiments. For instance, in the field of image processing, a false acceptance rate \citep[FAR, e.g.,][]{Jayadevan09} is measured by counting the frequency that DTW falsely recognize the true image, which is used to set the criteria for DTW applications. However, in astrophysical problems involving transients (such as the study in this work), we do not have the pre-knowledge about the central engine to define the ``correct behavior", and also do not have the luxury to let the source to repeat many times. As a result, an FAR cannot be defined and measured. The criteria stated above (e.g. $r >0.8$ \& $p <$ 0.003) may not be as robust as the FAR criteria used in other fields, but in any case, give a quantitative judgement about the similarities among different emission episodes.

\section{Results}

\subsection{GRB 140508A}

\begin{figure}
\begin{tabular}{c}
 \includegraphics[keepaspectratio,clip,height=4.5cm]{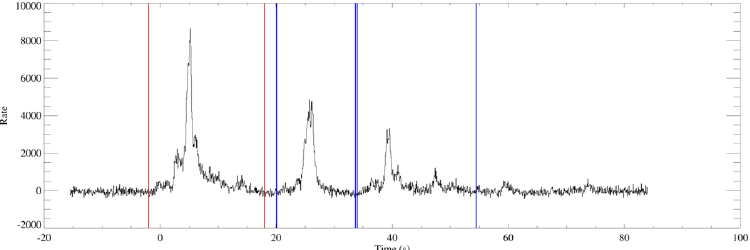} \\

\end{tabular}\\

\begin{tabular}{ccc}
\includegraphics[keepaspectratio,clip,width=0.32\textwidth]{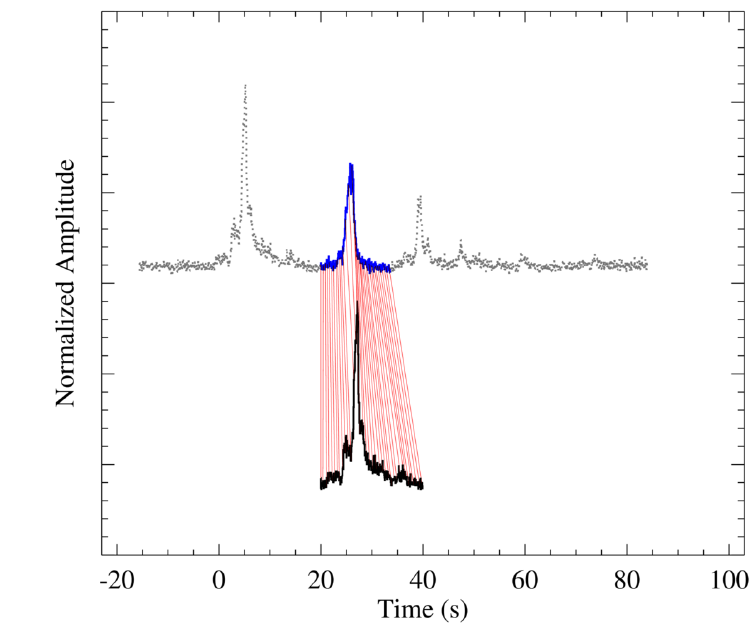} &\includegraphics[keepaspectratio,clip,width=0.23\textwidth]{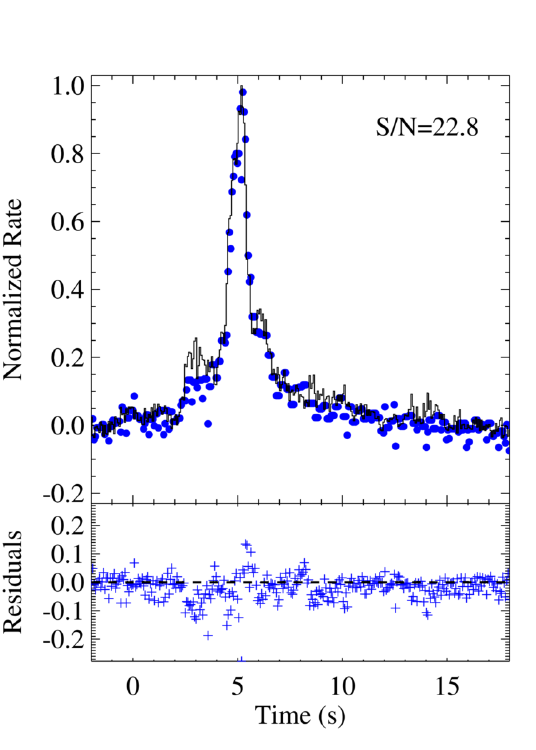} &
\includegraphics[keepaspectratio,clip,width=0.32\textwidth]{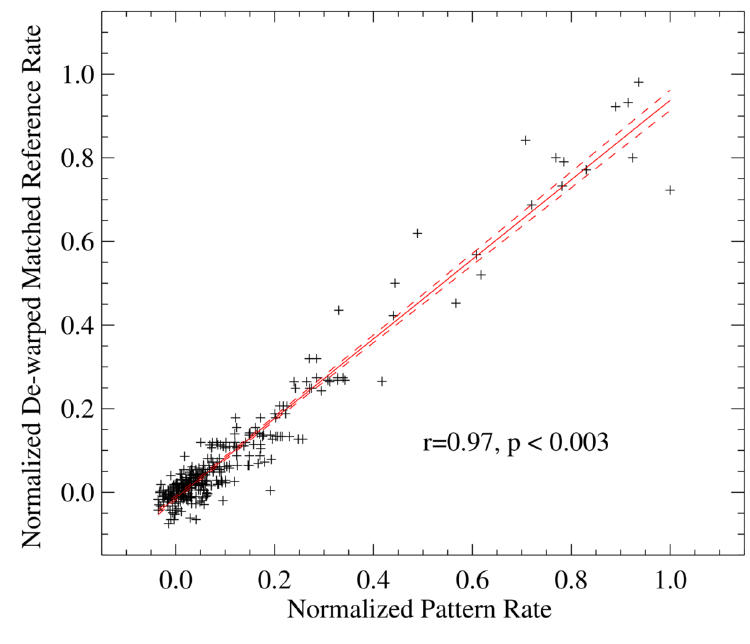}\\

\includegraphics[keepaspectratio,clip,width=0.32\textwidth]{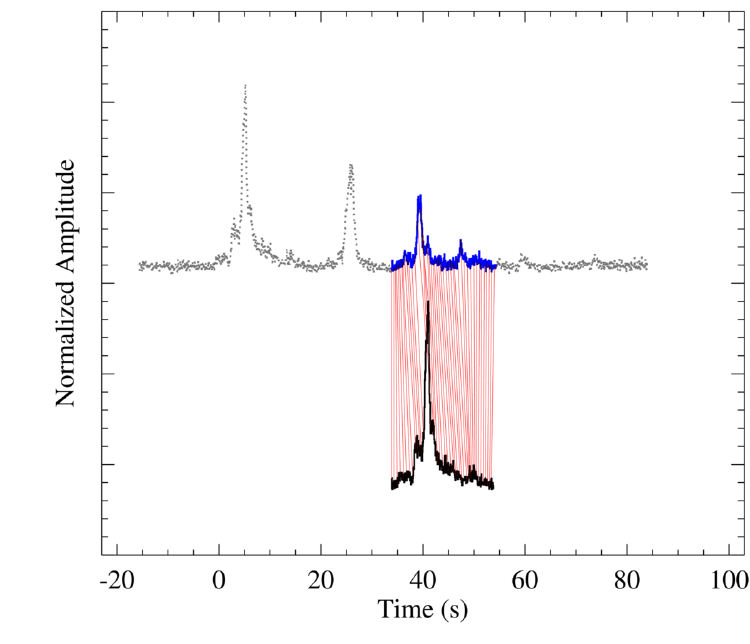} &\includegraphics[keepaspectratio,clip,width=0.23\textwidth]{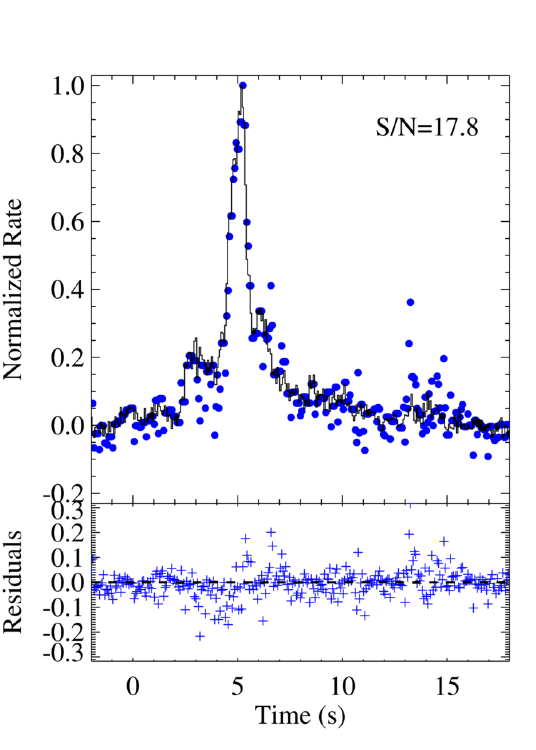}&
\includegraphics[keepaspectratio,clip,width=0.32\textwidth]{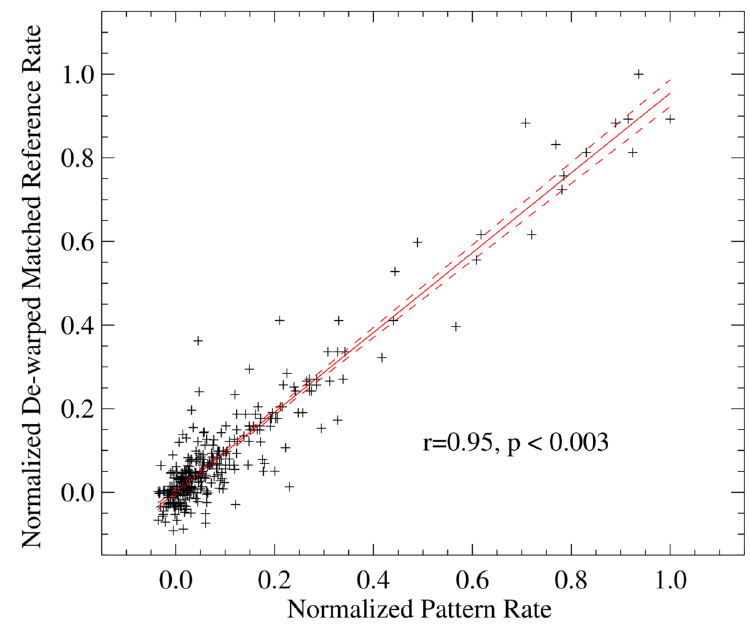}\\

\includegraphics[keepaspectratio,clip,width=0.32\textwidth]{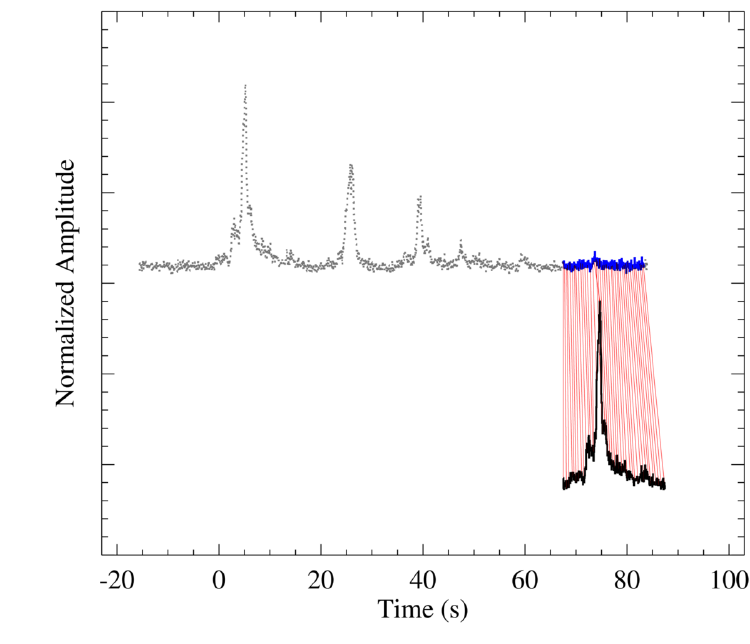} &\includegraphics[keepaspectratio,clip,width=0.23\textwidth]{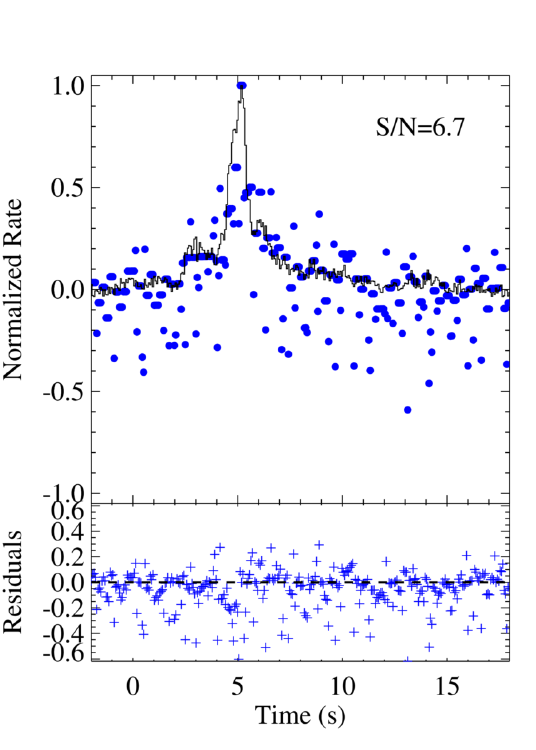}&
\includegraphics[keepaspectratio,clip,width=0.32\textwidth]{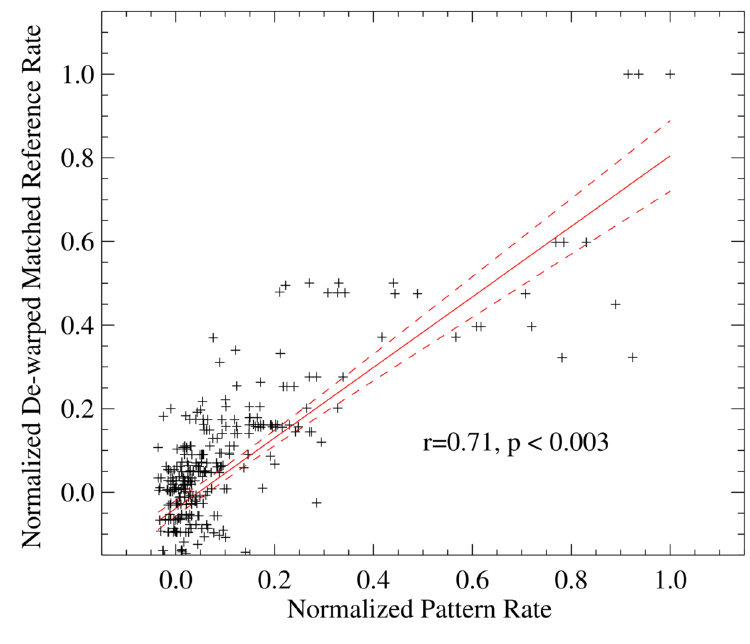}\\
\end{tabular}

\caption{ 
The case of GRB 140508A. {\it Top:} The original Fermi/GBM 10-800 keV light curve of GRB 140508A. The red vertical lines bracket the pattern region. The two pairs of blue vertical lines indicate the two successful matches. 
{\it Row 2-4:} Two accepted matches (Row 2 \& 3) and one rejected match (Row 4) of GRB 140508A. In each row, {\it Left} panel shows the DTW alignment between the pattern ({\it down}) and reference ({\it up}) light curves. Both light curves are normalized to unity. The entire reference light curve is plotted in grey while only the matched part is plot in blue. {\it Middle} panel is a comparison between the de-warped reference light curve ({\it blue dots}) and the pattern light curve ({\it black curve}) with the residuals plotted in the bottom. {\it Right} panel shows the correlation between the flux of the de-warped reference light curve and the pattern light curve, where the red solid lines mark the linear regression and the red dashed line marks the 95\% confidence region. The regression slope ($s$) and the Pearson's correlation coefficient ($r$) are marked in the figures.
}
\label{fig:case_grb}
\end{figure}

We first apply the method to GRB 140508A whose light curve displays 6 separated but similar emission episodes (Figure \ref{fig:case_grb}). Strong hard-to-soft spectral evolution is observed among those episodes. We select the first isolated episode emission above 3-$\sigma$ background level as the pattern, which is between $T_0$-2 s to $T_0$+18 s (where $T_0$ is the Fermi trigger time). Based on our criteria described above, we identified two successful matches, which are shown in Figures \ref{fig:case_grb}. For comparison, we also plot one rejected match (3rd row in Figure \ref{fig:case_grb}) which does not satisfy the successful criteria. The successful matches suggest that at least for this GRB, the central engine has a memory, which allowed it to reproduce a similar pattern of central engine activities for at least three times.

\subsection{GRB 151006A and its X-ray flare}

\begin{figure}
\begin{tabular}{ccc}
\includegraphics[keepaspectratio,clip,width=0.33\textwidth]{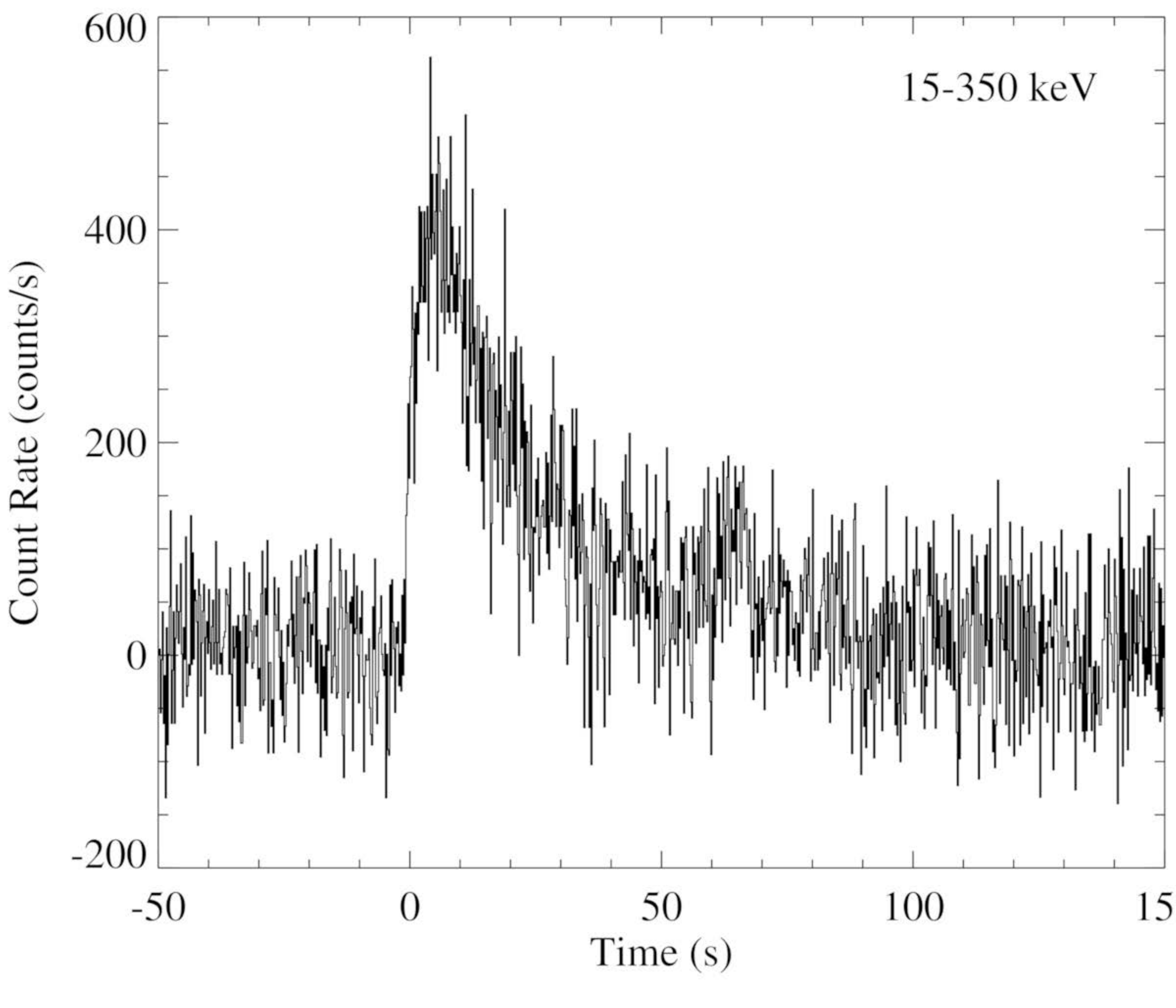} &
\includegraphics[keepaspectratio,clip,width=0.3\textwidth]{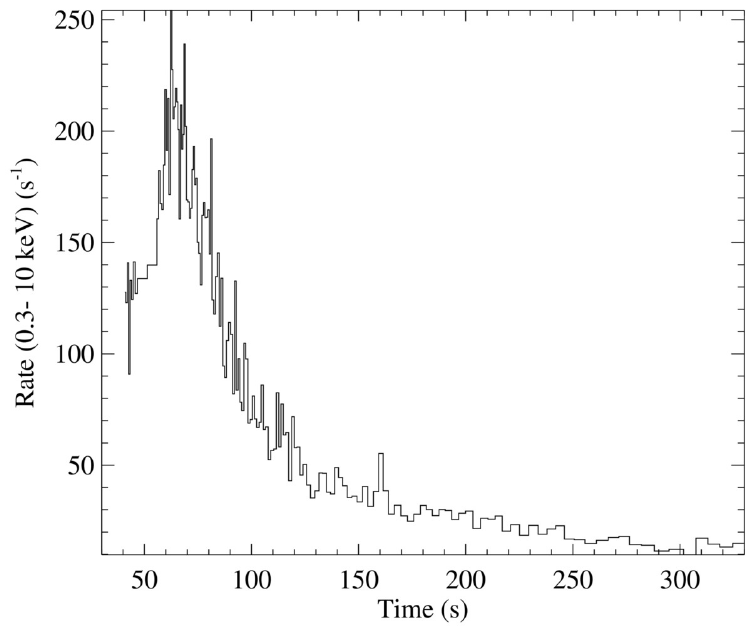} &
\includegraphics[keepaspectratio,clip,width=0.3\textwidth]{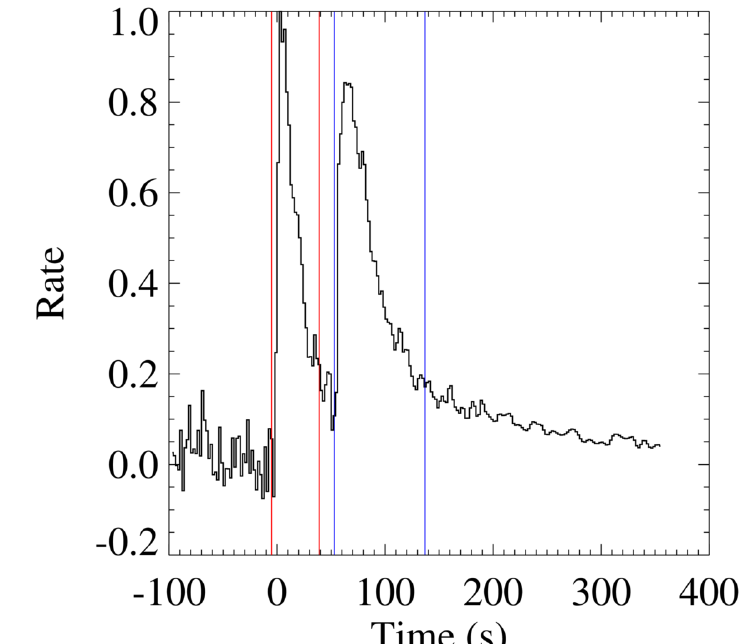} \\
\includegraphics[keepaspectratio,clip,width=0.33\textwidth]{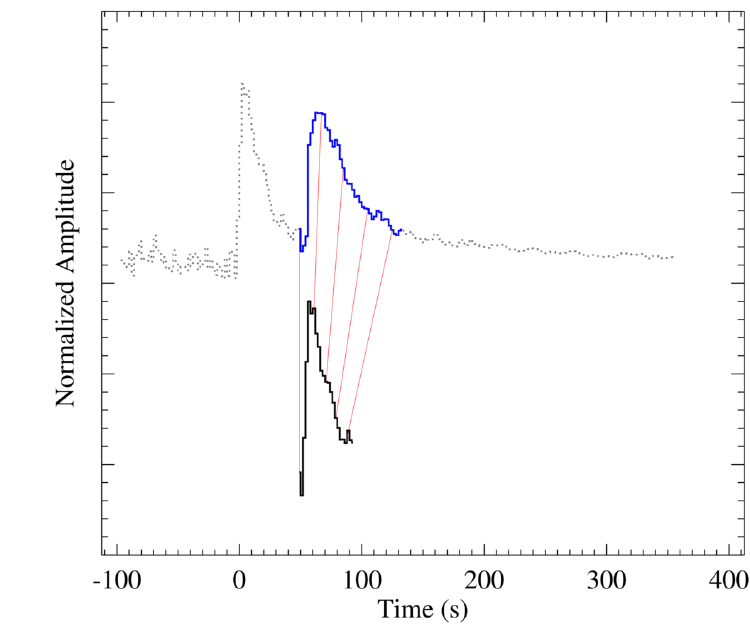} &\includegraphics[keepaspectratio,clip,width=0.3\textwidth]{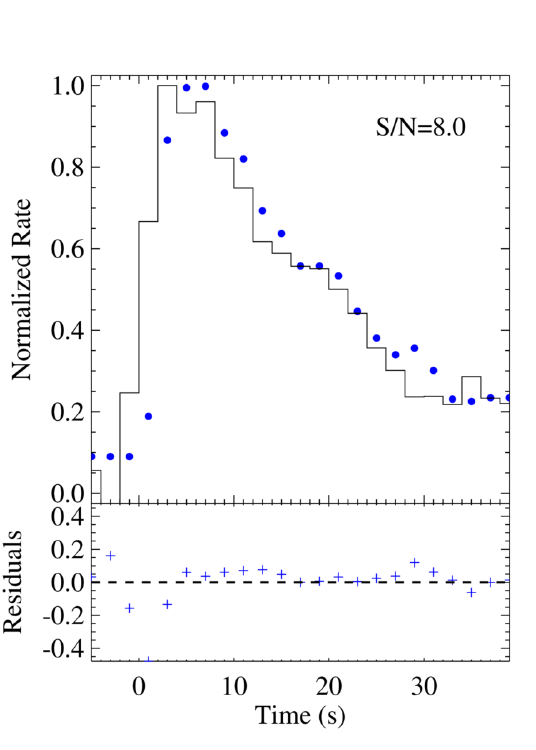} &
\includegraphics[keepaspectratio,clip,width=0.33\textwidth]{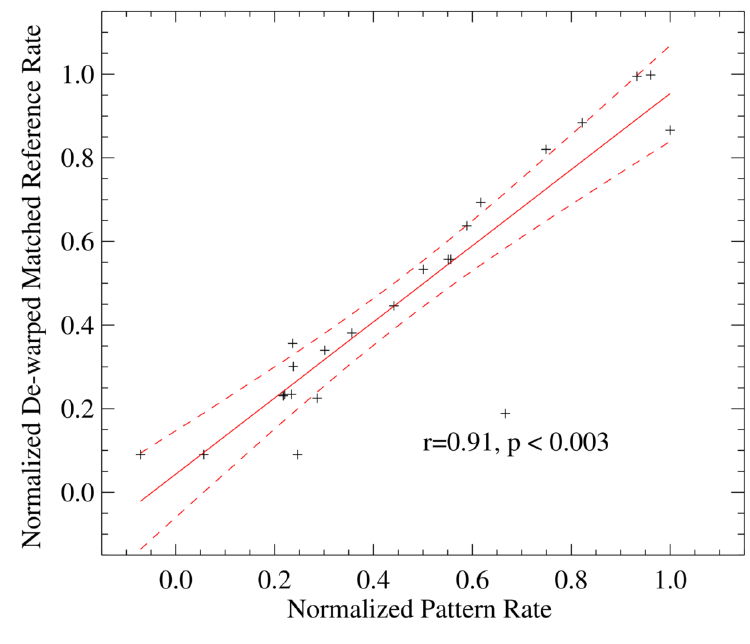}\\
\end{tabular}

\caption{The Case of GRB 151006A and its X-ray flare. {\it Top row}: The Fermi/GBM ({\it left, 15-350 keV}) and Swift/XRT ({\it middle, 0.3-10 keV}) light curves of 140508A, along with the combined ({\it right}) light curve to be used in our DTW method after binning into the same bin size. Pattern selection is bracketed with red vertical lines and the matched region is bracketed with blue vertical lines. In the combined light curve, the $\gamma$-ray light curve is normalized to unity, and the X-ray light curve is normalized artificially to $\sim$ 0.8 for clarity. Bottom row: The match between the prompt emission and X-ray flare emission in GRB 151006A. Legends and symbols are the same as Figure \ref{fig:case_grb}. } 
\label{fig:case_xray}
\end{figure}

To further investigate whether similar behavior appears in other GRBs as well, we look into the light curve data of other GRBs. Most GRBs have overlapping pulses (unlike GRB 140508A), which make it difficult to define a clean pattern to search for possible matches. We therefore include X-ray flares, which are the extension of the GRB central engine activities to the weaker and softer regime \citep{burrows05,zhang06,chincarini07,zhang14}, in the analysis. As an example, we show GRB 151006A, which has a single ``FRED" (fast-rise-exponential-decay, \citealt{2003ApJ...596..389K}) shape light curve with a duration about 20 s, and a significant X-ray flare between $\sim 50$ s and $\sim 100$ s (Figure \ref{fig:case_xray}). 
We take the 0-50 s prompt gamma-ray light curve as the pattern, and apply our DTW technique to search for the pattern in the X-ray flare light curve. 
Interestingly, we find a successful match (Figure \ref{fig:case_xray}) despite of the obviously different time scales and spectral ranges between the pattern and the reference light curves.

\subsection{Fermi/GBM Bursts 080823478 and 080823847 from SGR J0501$+$45165}
\begin{figure}
\begin{tabular}{cc}
\includegraphics[keepaspectratio,clip,width=0.3\textwidth]{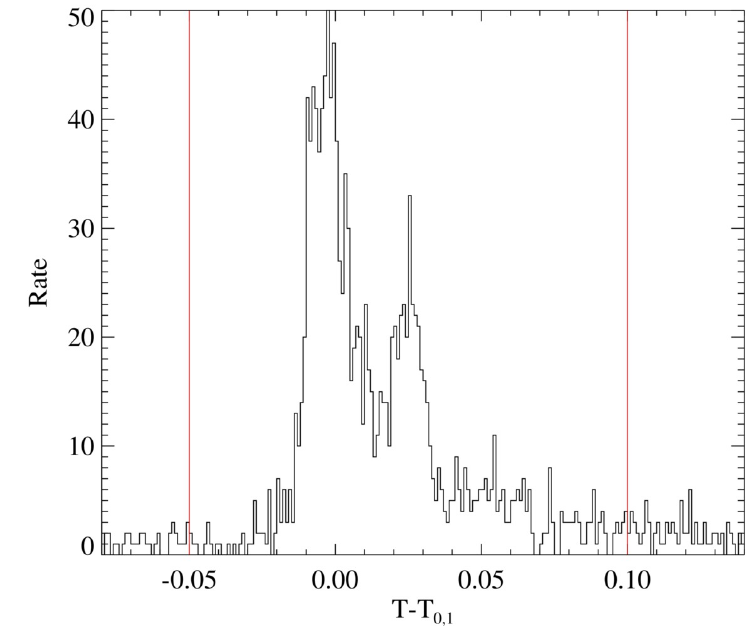} &
\includegraphics[keepaspectratio,clip,width=0.3\textwidth]{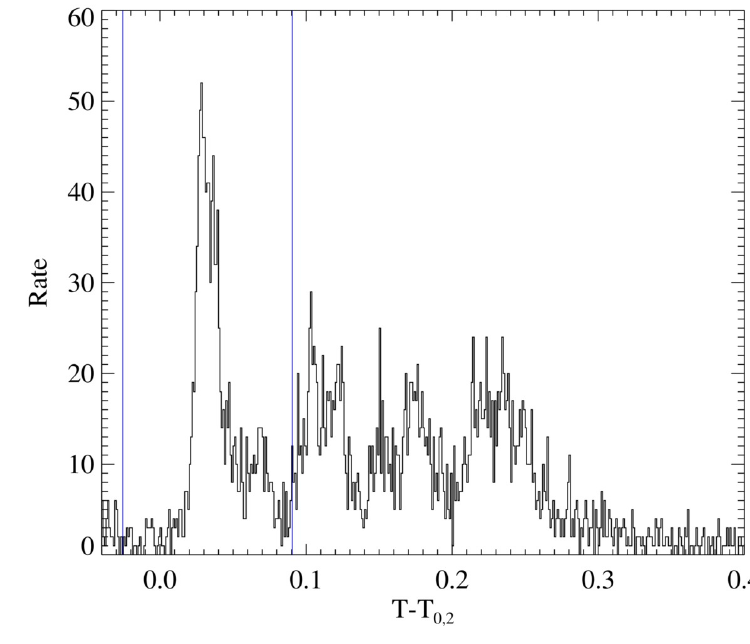}\\
\end{tabular}\\
\begin{tabular}{ccc}
\includegraphics[keepaspectratio,clip,width=0.36\textwidth]{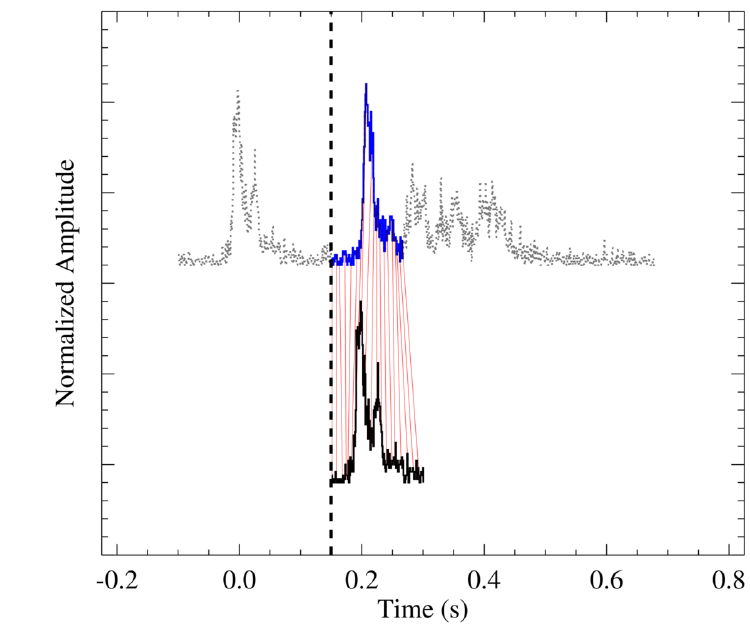} &\includegraphics[keepaspectratio,clip,width=0.3\textwidth]{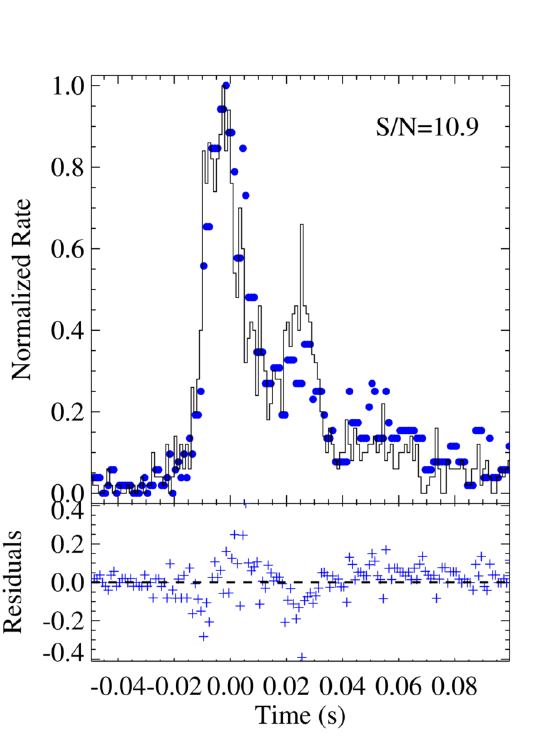}&
\includegraphics[keepaspectratio,clip,width=0.3\textwidth]{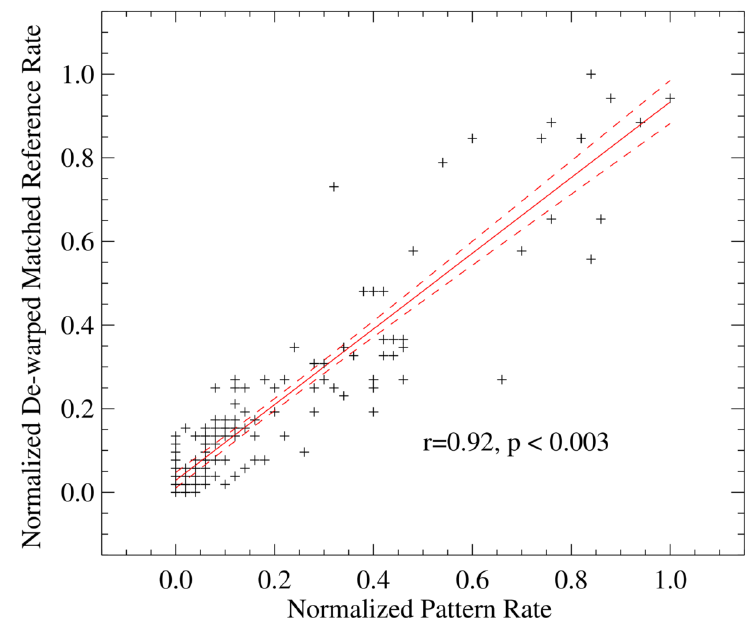}\\
\end{tabular}

\caption{The case for an SGR. {\it Top Row}: Light curves of Fermi/GBM bursts 080823478 ($T_{0,1}=$11:27:35 UT on Aug 08, 2008) and 080823847 ($T_{0,2}=$ 20:19:33 on Aug 08, 2008)) from SGR J0501 $+$ 4516. The red vertical lines bracket the pattern region. The blue lines mark the match. {\it Bottom Row}: A successful match between two flares in SGR J0501$+$45165. Legends and symbols are the same as Figure \ref{fig:case_grb}. Notice that in the lower level panel, the second flare is moved right after the first flare for direct comparison, even if they are separated by about 9 hours. Such 9-hour separation is marked with a dashed line. } 
\label{fig:case_sgr}
\end{figure}

 In order to investigate whether the central engine memory may also apply to other transient events, we apply our DTW method to the light curves of SGR bursts, which usually happen sporadically, with separations ranging from hours to months. 
We applied our method to a pair of bursts, named as 080823478 and 080823847, from SGR J0501$+$45165 \citep{2013ApJ...778..105L}. These two bursts happened in the same day but were separated by about 9 hours. Their light curves are shown in Figure \ref{fig:case_sgr}. Interestingly, we find one successful match (Figure \ref{fig:case_sgr}) between the first flare and the first emission episode of the second flare. 
Despite the previous failure to identify a QPO from the light curves of this source \citep{2013ApJ...768...87H}, we show that the magnetar central engine of SGR J0501$+$45165 seems to also carry a memory of its activities.

\section{Summary and Discussions}

{ By applying the Dynamic Time Warping (DTW) method for the first time to astronomical objects, we found that similar light curve patterns do exist within individual bursts for at least some GRBs. We also found that such a similarity exists between the light curves of a GRB and its X-ray flare as well as distinct emission episodes in the flaring emission in SGRs, in particular, in SGR J0501+45165.

 Following physical processes may leave imprints in the light curve of a GRB: 1. the erratic central engine activity; 2. processes during jet propagation (e.g. jet modulated by a stellar envelope, e.g. \citealt{morsony10}); 3. rapid variability in the emission region due to magnetic reconnection or turbulence \citep{narayan09,zhangyan11,zhangzhang14}; and 4. geometric effects (e.g. due to jet precession, \citealt{lei07}). The mechanisms 2 \& 3 invoke instabilities, whose behaviors are erratic, likely not memory-like physical. The variability of a precessing jet should be periodical, which is not true for the cases we have studied. This makes central engine itself as a possible source of memory. There are two widely discussed types of central engines for GRBs: a hyper-accreting black hole (BH) and a rapidly spinning, millisecond pulsar (or magnetar) \citep[][for a recent review]{kumarzhang15}. For a BH engine model, the variability may originate from a variation in the accretion rate, which is in turn related to the mass-feeding rate. It is hard to imagine how a memory-like process could regulate the accretion rate in certain patterns. On the other hand, a magnetar central engine may have the potential to induce memory-like behaviors. For example, \cite{kluzniak98} invoked the disruption of magnetic loops formed in a differentially rotating neutron star central engine to interpret variability of GRB prompt emission. A similar mechanism applied to a massive neutron star after double neutron star mergers could interpret X-ray flares following short GRBs \citep{dai06}. For these processes, similar to the solar cycle activities, the mechanism to produce an emission episode invokes similar physical processes, including winding up magnetic field lines due to differential rotation, break field lines at a critical condition, and release high energy particles and emission abruptly. As the emission episode is over, the same process would repeat itself and generate a similar emission episode\footnote{The similar process may apply to a BH central engine, if the jet is launched from a highly magnetized disk rather than BH itself \citep{yuanzhang12}.}. Unlike the Sun whose differential rotation rate is essentially constant for many cycles (and hence, produce a 11-year periodic behavior), GRBs invoke a rapidly evolving central engine with the power quickly decrease with time. One therefore does not expect a strict periodicity, but expect the late-time emission episodes be weaker and more temporally stretched with respect to the early-time ones. This would give a possible interpretation to the discovery in this paper. Since SGRs are Galactic magnetars, the discovery of memory patterns in these objects is consistent with this interpretation. Since GRB magnetar models produce highly magnetized winds, the results also lend supports to the GRB emission models that invoke dissipation of Poynting flux energy as the resource of GRB prompt emission \citep{zhangyan11}.}

\acknowledgments {We acknowledge the pubic DTW code in R language \citep{Rcore15,Toni09,Paolo08} and the anonymous referee for helpful comments.} BBZ thank Peter Veres, Lin Lin, Sam Oates, Valerie Connaughton, Michael S. Briggs, Wei-Hua Lei, Wei Chen, Juan Carlos-Tello, En-Wei Liang, Guillaume Belanger, George Younes, Robert Preece, David N. Burrows, Neil Gehrels, J. Hakkila, Kohta Murase, Lang Shao and Yuan-Chuan Zou for helpful suggestions and discussions on this investigation. BZ acknowledges partially support by NASA through NNX14AF85G and NNX15AK85G. BBZ and AJCT acknowledge support from the Spanish Ministry Projects AYA 2012-39727-C03-01 and AYA2015-71718-R. The computation of this work were performed on BBZ's personal computational system - the Scientist Support Computational System (SSCS). Part of this work used BBZ's personal IDL code library ZBBIDL and personal Python library ZBBPY.

\appendix

\section{Basics of The Dynamic Time Warping Method and Its Applciaiton to Astronomical Light Curves}

Dynamic time warping (DTW) was originally developed for speech recognition\citep{Sakoe78,Rabiner83,Juang91}, i.e. to find an optimal
alignment between two given time series under certain restrictions. The method allows one sequence to be warped in a nonlinear manner to match the other. For example, similarities in speech patterns (e.g., same words or phases) can be detected using DTW, even if one person was speaking faster than the other, or if there were accelerations/decelerations during the course of an observation. This technique is now widely used in many other fields as a tool of data mining and information retrieval, such as in computer vision science \citep[e.g., signature/fingerprint recognition;][]{Munich99,Kovacs2000}, bioinformatics\citep[e.g.,][]{aach01}, chemical engineering\citep[e.g.,][]{Dai11}, to automatically handle the time deformations and different speeds associated with time-dependent data. We refer the interested readers to Keogh \& Ratanamahatana (2005) for a more detailed technical treatment. 

\begin{figure*}[b]
\includegraphics[keepaspectratio,clip,width=0.5\textwidth]{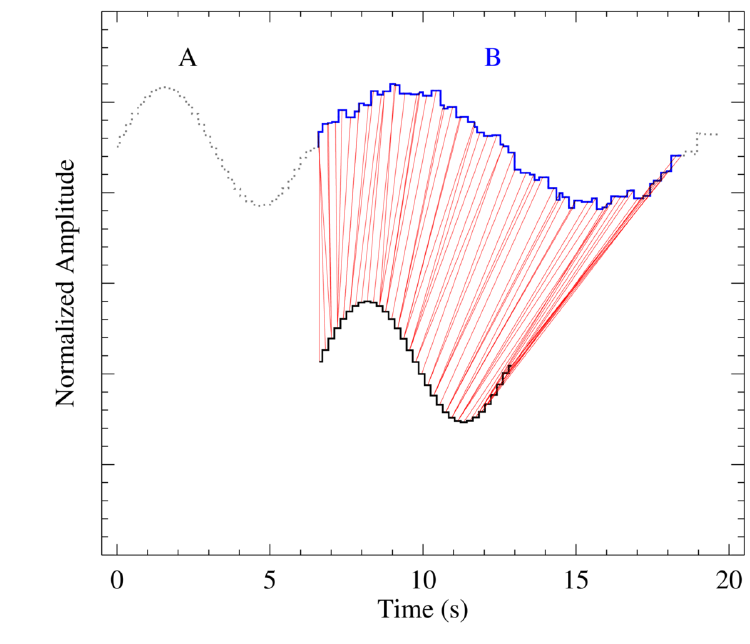}

\caption{ 
A toy example illustrating how the DTW method works for light curves. An astronomical object is assumed to emit its light curve with a perfect sine function in Episode A, and then emits a distorted sine shape light curve (with fluctuations in amplitude added), and slows down in time {\it non-uniformly} by an averaged factor of 2, in Episode B. The red lines, calculated by the DTW method, show the alignments between the two episodes. }
\label{fig:A1}

\end{figure*}

A toy example is presented in Figure \ref{fig:A1} to illustrate the method and its application to astronomical light curves. In this example, we assumes that an astronomical object emits its light curve with a perfect sine function in Episode A, then emits a distorted sine shape light curve (with fluctuations in amplitude included), and slows down in time {\it non-uniformly} by an averaged factor of 2 in Episode B. By applying the DTW method, one can successfully find the similarity between the light curves in the two episodes (see point-to-point matches through red lines).

\end{document}